\documentstyle[aabib,epsfig]{l-aa-ps}
    
    \newcommand{\E}{{\em Einstein}}

    \newcommand{\cq}{\mbox{$ \chi^2$}}

    \newcommand{\feh}{\mbox{[Fe/H]}}

    \newcommand{\dof}{\mbox{degrees of freedom}}

    \newcommand{\bcet}{\mbox{$\beta$~Cet}}
\begin{document}
\thesaurus{6(08.09.2 Capella; 08.12.1; 08.01.2; 13.25.5)}

\title{A SAX/LECS X-ray observation of the active binary
  Capella}

\author{F. Favata\inst{1} \and R. Mewe\inst{2} \and N.~S.
  Brickhouse\inst{3} \and R. Pallavicini\inst{4} \and G.
  Micela\inst{4} \and A.~K. Dupree\inst{3}
}

\institute{Astrophysics Division -- Space Science Department of ESA, ESTEC,
 Postbus 299, NL-2200 AG Noordwijk, The Netherlands
\and
SRON Laboratory for Space Research, Sorbonnelaan 2, NL-3584, Utrecht,
 The Netherlands 
\and
Harvard-Smithsonian Center for Astrophysics, 60 Garden St., 02138
 Cambridge, Mass., USA
\and
 Osservatorio Astronomico di Palermo, 
 Piazza del Parlamento 1, I-90134 Palermo, Italy 
\\
}

\offprints{F. Favata (fabio.favata@astro.estec.esa.nl)}

\date{Received date ; accepted date}

\maketitle 
\begin{abstract}

  We present a SAX/LECS X-ray observation of the active binary
  Capella, the first coronal source observed in the SAX Guest
  Investigator program. The analysis of this observation, performed
  using the {\sc mekal} plasma emission code, shows that the LECS
  spectrum is well fit by a two-component optically-thin plasma model.
  A differential emission measure (DEM) obtained by direct inversion
  of the X-ray spectrum shows no additional features in addition to
  the double-peaked structure implied by the direct two-temperature
  analysis.  Such a simple temperature stratification is however not
  compatible with the EUVE emission from the same object, which is
  well represented by a more complex DEM, with a power-law-like tail
  toward the low temperatures. At the same time, the EUVE-derived DEM
  predicts well the softer part of the Capella LECS spectrum, but it
  fails to correctly reproduce the higher energy part of the Capella
  LECS spectrum. Possible causes for this discrepancy are discussed.
  The coronal metallicity derived from the SAX observation is
  compatible both with the EUVE-derived metallicity as well as with
  the photospheric metallicity of Capella, thus showing no evidence
  for coronal under-abundances.


\end{abstract}

\section{Introduction}
\label{sec:intro}

The thermal nature of the emission from the coronae of late-type stars
as a class was established through the study of the large number of
low-resolution X-ray spectra taken with the \E\ IPC, which were well
fit by a thermal spectrum, requiring either one or two discrete
temperature components (e.g. \cite{scs+90}). While the limited
spectral resolution of the IPC did not allow to investigate the
presence of a more complex temperature structure in the emitting
plasma, analysis of the emission from the solar corona showed that
more complex structure is likely to be present in coronal plasmas. The
newer instruments available for X-ray and EUV spectroscopy, with their
improved spectral resolution and energy coverage, have all added new
complexity to the picture of stellar coronal emission. The data from
the ASCA/SIS detector have been shown to often require deviations in
the plasma metal abundance from solar values (although the issue of
deviations from stellar photospheric abundances, physically more
relevant, has not been thoroughly investigated), leading to a debate
about supposed widespread under-abundances in coronal plasmas (the
so-called Metal Abundance Deficiency syndrome, or MAD).  Still, the ASCA data of
coronal sources have mostly been modeled with two discrete temperature
components.

The data from the EUVE spectrographs, with their much higher
resolution which allows for individual lines from the various Fe
ionization states to be well resolved, have been shown to require a
more complex temperature structure for their modeling, with two
discrete temperature components in general not supplying a
satisfactory description of the data. The EUVE-derived differential
emission measure (DEM) of many active binaries for example shows a
characteristic feature at $\simeq 0.5$\,keV often referred to as a
``bump'' (\cite{dbh96}).

The Low Energy Concentrator Spectrometer (LECS, \cite{pmb+97}) on-board
the SAX satellite (\cite{bbp+97}) opens a new window for the study of
the X-ray emission of coronal sources, thanks to its wide spectral
coverage (from 0.1 to 10\,keV) and good spectral resolution, in
particular at the low energies, where it has a resolution comparable
to CCD detectors. Energies below $\simeq 0.5$\,keV are not covered by
the ASCA/SIS CCD detector, yet it is precisely at these energies that
typical coronal sources emit the largest photon flux. The LECS
bandpass thus fills the gap between the EUVE energy coverage and the
ASCA one.

Capella is a well-known nearby bright multiple system, with a slowly
rotating G8\,III primary and a fast rotating F8\,III secondary (which
has been identified as the site of strong UV chromospheric emission,
\cite{lwj+95}), and a long orbital period ($\simeq 104$\,d).  Its
photospheric metallicity is close to solar, with some evidence for
mild under-abundance, with \cite*{ps92} reporting a Ca abundance
compatible with the solar value, and \cite*{mcw90} reporting $\feh =
-0.37 \pm 0.22$ (equivalent to 0.43 times solar, with a $1\,\sigma$
range of 0.26--0.71). Thanks to its high X-ray luminosity and to its
small distance from the Earth, Capella is a well studied coronal
source, having been observed with essentially all the soft X-ray and
UV/EUV detectors flown to date. It is also the first coronal source to
have been observed in the first year of the SAX guest observing
program.

The corona and transition region of Capella has been extensively
studied with the EUVE spectrographs, and a detailed analysis of the
EUV spectrum (\cite{dbd+93}; \cite{bri96}) shows that the differential
emission measure distribution has a ``bump'' around $\simeq 0.5$\,keV.
The EUV spectrum, and in particular the line to continuum ratio, shows
that the Fe abundance ($0.88\pm 0.13$ times solar) is compatible with
the solar photospheric value, and and close to the upper bound of the
reported photospheric abundance, as are the abundances of the other
species whose lines are visible in the EUV spectrum.

\section{Observations and data reduction}

The SAX observation of Capella took place on Oct. 4 and 5, 1996 but
only 12\,ks of observation were obtained with the LECS detector,
rather than the 30\,ks which had been allocated. The instrument
performed nominally, and no evidence for flaring activity or
significant time variability is present in the light curve of the
source.

\begin{figure}[thbp]
  \begin{center}
    \leavevmode
    \epsfig{file=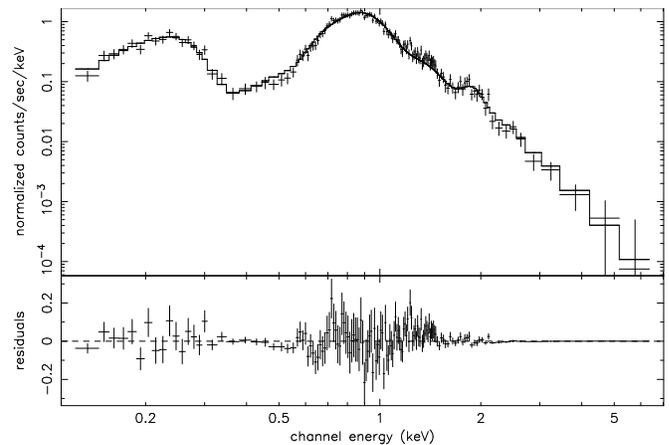, width=9.cm, bbllx=5pt, bblly=30pt,
    bburx=687pt, bbury=473pt, clip=}
  \end{center}
  \caption{The observed SAX/LECS spectrum of Capella, together with the
    best-fit two-temperature {\sc mekal} spectrum and the fit
    residuals (expressed in terms of their contribution to the .}
  \label{fig:best}
\end{figure}

The data were reduced through the LECS pipeline software (SAX-LEDAS
V.\,1.4.0), the extraction of the source and background spectra was
done within the XSELECT package, the response matrix was produced with
release 3.2.0 of the LEMAT software, and the spectral analysis was
done using the XSPEC V.\,9.0 and SPEX V.\,1.10 packages.  The source
spectrum was extracted from a circle 35 pixels in radius
(corresponding to 8\,arcmin), while the background was derived from
long observations of empty fields, and extracted from the same area as
the source. The extraction radius of 8\,arcmin is the default in the
SAX-LEDAS software pipeline, and has been chosen to allow for 95\% of
the source counts at 1\,keV to be included. The source spectrum was
re-binned so to have at least 20 counts per re-binned channel, and
channels with energies below 0.1 and above 6.0\,keV were discarded.
The resulting source count rate is 0.78\,cts\,s$^{-1}$, to be compared
with a background count rate, for the same region and spectral
interval, of 0.011\,cts\,s$^{-1}$. The (background-subtracted)
spectrum is shown in Fig.~\ref{fig:best} (together with the best-fit
two-temperature model discussed below).

\section{Results}

The first step in our analysis of the LECS Capella spectrum has been
to fit it with an optically-thin plasma emission model with two
discrete temperature components. The {\sc mekal} plasma emission model
(\cite{mkl95}) was used throughout, as implemented in XSPEC 9.0. The
column density toward the source was fixed to $1.8\times
10^{18}$\,cm$^{-2}$, the value derived from the analysis of the HST
Ly$\alpha$ data (\cite{lbg+93}).
While (as shown by \cite{fmp+97}) the LECS is rather
sensitive to global abundance variations, its limited spectral
resolution makes the determination of coronal abundances of individual
elements much less reliable, especially at signal-to-noise ratio of
the spectrum discussed here. Indeed, given the satisfactory reduced
\cq\ of the fit with a model in which only the global abundance is
left free to vary, the variation of individual abundances is not
necessary to fit the present spectrum. Thus, in the present paper,
only global abundance variations are discussed. The resulting best-fit
model for a two-temperature {\sc mekal} model is shown, together with
the observed spectrum and the fit residuals, in Fig.~\ref{fig:best}.
The resulting reduced \cq\ is 1.04 (with 138 \dof), 
resulting in a fully satisfactory fit. The best-fit temperatures are
$0.66\pm 0.05$ and $1.04\pm0.2$\,keV, the emission measures are
5.73$\pm 1.0\times 10^{52}$ and 1.49$\pm 1.1 \times
10^{52}$\,cm$^{-3}$ for the cool and the hot component respectively
(assuming a distance of 13.4\,pc; note the much larger uncertainty on
the parameters of the hot component), and the best-fit coronal
metallicity is $0.68\pm 0.05$ times the solar photospheric value, a
value compatible with the photospheric Fe abundance of \cite*{mcw90}.

A direct inversion performed using the ``multi-temperature'' algorithm
contained in the SRON SPEX package yields a two-component DEM (shown in
Fig.~\ref{fig:dem} by the dashed line) very similar to the one implied
by the two-temperature analysis, with two peaks and very little if any
additional structure. To produce this DEM, the global metallicity was
fixed to the best-fit value derived from the two-temperature analysis.

\subsection{Comparison with the EUVE-derived DEM}

\begin{figure}[thbp]
  \begin{center}
    \leavevmode
    \epsfig{file=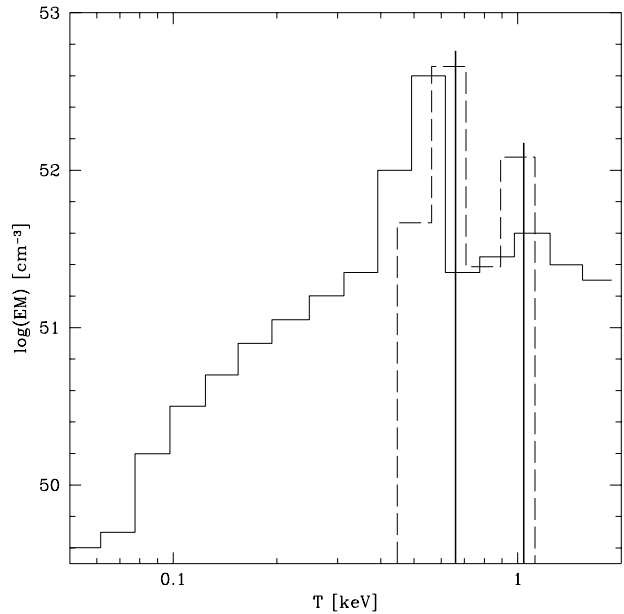, width=8.8cm, bbllx=0pt, bblly=150pt,
      bburx=580pt, bbury=700pt, clip=}
    \end{center}
  \caption{The  DEM implied by the two-temperature analysis is plotted
    (two thick vertical lines) together with the DEM obtained by
    direct inversion of the LECS spectrum (dashed-line histogram) and
    the EUVE-derived DEM of Brickhouse (1996) (continuous-line
    histogram).}
  \label{fig:dem}
\end{figure}

The EUVE spectrum of Capella has been discussed by \cite*{dbd+93},
\cite*{smo+95} and \cite*{bri96}. Their analysis shows the presence of
a complex structure in the DEM, with a minimum around 0.1\,keV, rising
toward higher temperatures, and with a fairly narrow feature (the
``bump'') around 0.4--0.5\,keV. Substantial emission measure appears
to be present up to $\simeq 1.5$\,keV. \cite*{smo+95} and
\cite*{mos+96} have compared and discussed in detail the results for
the DEM derived from different EUVE analyses and from observations
with various other instruments.  The EUVE-derived DEM of \cite*{bri96}
is shown in Fig.~\ref{fig:dem} by the continuous line histogram (note
the logarithmic scale).

A synthetic spectrum computed using the EUVE-derived DEM produces a
very good fit to the soft component ($E \la 0.5$\,keV) of the LECS
spectrum, both for the spectral shape and for the normalization, with
an acceptable resulting \cq. However, the same DEM does not do a good
job at reproducing the $\simeq 1$\,keV peak of the spectrum and the
hard tail. The fit improves if the relative normalization of the
hotter ``plateau'' of the EUVE-derived DEM is left free to vary. The
normalization of the high-T plateau becomes in this case much higher,
requiring values which are incompatible with the EUVE line fluxes. In
particular, several lines from Fe\,{\sc xviii}, Fe\,{\sc xix} and
Fe\,{\sc xx} become over-predicted when the additional
high-temperature emission measure is included, since their emissivity
curves peak in the range between $\simeq 0.5$ and $\simeq 1$\,keV
(\cite{brs95}) and their intensity results from both from the ``bump''
in the DEM ($\simeq 0.5$\, keV) as well as from the $\simeq 1$\,keV
component.  Thus, the presence of a large emission measure at $\simeq
1$\,keV, required by the LECS data, produces large imbalances in the
predicted EUVE line flux.

A better fit to the LECS spectrum can be achieved by leaving all the
hotter components of the EUVE-derived DEM free to vary. In this case
the fitting process ``collapses'' the flat high-temperature
distribution resulting from the analysis of the EUVE line fluxes onto
a single component, whose normalization again needs, to properly fit
the LECS spectrum, to be much higher than allowed by the EUVE line
fluxes. While the presence of a sharp (and with higher normalization)
high-temperature feature in the DEM improves the fit to the LECS data,
it is again not compatible with the EUVE spectrum. Thus, it appears
that a single DEM cannot satisfactorily fit both the EUVE spectrum as
analyzed by \cite*{bri96} and the LECS spectrum.

Given that the LECS and EUVE observations are not simultaneous,
intrinsic source variability may play a role. The EUVE spectrum from
which the DEM used here has been derived is actually the sum of a
number of individual spectra taken at different times, ensuring
maximal signal to noise. Analysis of the DEM derived from individual
spectra shows that while the variability of the ``bump'' component is
limited (i.e. $\la 30$\%), the higher-temperature emission components
tend to show a much higher variability, up to a factor of 4
(\cite{db96}). Higher variability for the hotter component was seen in
many active binaries, using \E\ SSS data, by \cite*{swh+81}, although
Capella was the only object, in their data, showing {\em no}
variability in its hot component. Thus, part of the discrepancy
between the ratio of high- and low-temperature emission measure
necessary to fit the EUVE and SAX spectra could be due to intrinsic
variability of the emission from the higher-temperature plasma.
Simultaneous observations will be needed to ascertain whether source
variability alone can explain the observed discrepancy.

\subsection{Comparison with other previous observations}

A review of the existing X-ray data on Capella shows a rather puzzling
variety of results. While most of the spectra have been fit with
two-temperature models, the best-fit parameters vary widely with
regard to both temperatures and emission measure ratios.
Two-temperature and DEM fitting of the spectra taken with the various
spectrometers on board \E\ (transmission grating and solid state,
\cite{mgw+82}, \cite{swh+81}) and on the EXOSAT transmission grating
(\cite{lms+89}) have revealed a structure with two temperature maxima,
concentrated around 0.3--0.5 and 1--3\,keV, respectively.

The \E\ IPC spectrum of Capella, when fit with a 2-temperature model
yielded best-fit temperatures of 0.21 and 0.69\,keV and a ratio
between the cool and hot emission measure of 0.15. The \cq\ of the fit
was, at 2.3, rather poor (\cite{scs+90}). The ROSAT PSPC data yielded
best-fit temperatures of 0.2 and 1.0\,keV and an emission-measure
ratio of 0.33, with an acceptable \cq\ (\cite{dlf+93}). The EXOSAT TGS
data yelded a rather hot spectrum, with temperatures of 0.43 and
2.16\,keV, and an emission-meaure ratio of 0.34 (\cite{lms+89}). A
re-analysis of the EXOSAT ME data (\cite{opt97}) also shows similar
temperatures, at 0.56 and 2.2\,keV, and an inversion of the
emission-measure ratio, at 3.2.  Finally, \cite*{bb96} report a 2-T
fit with variable abundance to the PSPC data, with temperatures of
0.13 and 0.68\,keV, a very low emission-measure ratio of 0.026 and a
range of acceptable metallicities of 0.15--0.3 times solar. The \cq\ 
is however, at 3.4, rather poor.

These fits are not necessarily directly comparable since they are
based on different plasma emission codes, use different fit acceptance
criteria, and make different assumptions about the model metallicity
(either kept fixed to the solar value or left free to vary). Moreover,
temporal variability between the various observations is also likely
to play a role.

The only published results on the ASCA SIS data of Capella are the
preliminary 2-T fits of \cite*{dra96} and \cite*{whi96}, which were
made will all the individual abundances left free to vary, and
apparently did not converge to a satisfactory fit so that their
best-fit parameters (which would imply a factor 0.2-0.3 under-abundance
of metals with the respect to the solar values) cannot directly
compared with the results of our fits to the LECS spectra (which were
made with only the global metallicity left as a free parameter and
which did converge to a satisfactory fit). A comparison of the LECS
data with the ASCA SIS data, using both available and still to be
released ASCA archival data, is deferred to a future paper.

\section{Conclusions}

The SAX LECS spectrum of Capella is well fit by a simple
two-temperature spectral model using the {\sc mekal} plasma emission
code. A DEM derived by inversion of the LECS spectrum yields a similar
temperature structure.  Such a simple 2-T model however does not
provide a good fit to a (non-simultaneous) observation by EUVE, and
thus is likely not a good description of the full temperature
structure of the corona of Capella. The LECS spectrum is compatible
with the broad low-temperature peak present in the EUVE-derived DEM,
but indicates a much larger emission measure at a temperature greater
than $10^7$\,K.  The ratio between the emission measures of the cool
and of the hot component required to fit the LECS spectrum is
incompatible with the ratio of the emission measure at the same
temperatures required to fit the EUVE spectrum. The coronal
metallicity derived from our fits to the LECS spectrum is compatible
both with the EUVE-derived data as well as with the published values
of photospheric abundance. Capella therefore (similarly to \bcet,
\cite{mfp+97}) show no evidence for a significant ``Metal Abundance
Deficiency'', contrary to early reports based on both ROSAT PSPC data
(\cite{bb96}) and a preliminary analysis of ASCA SIS spectra
(\cite{dra96}; \cite{whi96}).

\acknowledgements{We would like to thank A.~Maggio for the many useful
  discussions and suggestions. The BeppoSAX satellite is a joint
  Italian and Dutch program. The Space Research Organization of the
  Netherlands (SRON) is supported financially by NWO, the Netherlands
  Organization for Scientific Research. R. P. and G. M. acknowledge
  financial support from GNA-CNR, ASI and MURST.}

\end{document}